\documentclass[%
 reprint,
superscriptaddress,
 amsmath,amssymb,
 aps,prl,floatfix
]{revtex4-1}

\usepackage{todonotes}
\usepackage{graphicx}%
\usepackage{dcolumn}%
\usepackage{bm}%
\usepackage[normalem]{ulem}
\usepackage{microtype}
\usepackage{xcolor}
\usepackage{amsmath,amssymb,amsfonts,mathtools,cancel}
\usepackage{mhchem}

\usepackage[a4paper, top=2cm, bottom=2cm, left=2cm, right=2cm, marginparwidth=3.3cm, marginparsep=2mm]{geometry}

\newcommand{\bra}[1]{\ensuremath{\left \langle #1 \right \vert}}
\newcommand{\ket}[1]{\ensuremath{\vert #1 \rangle}}

\newcommand{\brckt}[3]{\ensuremath{\left \langle #1 \left| #2 \right| #3 \right \rangle}}

\begin{document}

\title{Nuclear quantum effects enter the mainstream}

\author{Thomas E. Markland}
\email{tmarkland@stanford.edu}
\affiliation{Department of Chemistry, Stanford University, Stanford, California 94305, USA}

\author{Michele Ceriotti}
\email{michele.ceriotti@epfl.ch}
\affiliation{Laboratory of Computational Science and Modeling, Institute of Materials, \'Ecole Polytechnique F\'ed\'erale de Lausanne, 1015 Lausanne, Switzerland}

\date{\today}

\begin{abstract}
Over the past decades, atomistic simulations of chemical, biological and materials systems have become increasingly precise and predictive thanks to the development of accurate and efficient techniques that describe the quantum mechanical behavior of electrons. However, the overwhelming majority of such simulations still assume that the nuclei behave as classical particles. While historically this approximation could sometimes be justified due to complexity and computational overhead, the lack of nuclear quantum effects has become one of the biggest sources of error when systems containing light atoms are treated using current state-of-the-art descriptions of chemical interactions. Over the past decade, this realization has spurred a series of methodological advances that have led to dramatic reductions in the cost of including these important physical effects in the structure and dynamics of chemical systems. Here we show how these developments are now allowing nuclear quantum effects to become a mainstream feature of molecular simulations. These advances have led to new insights into chemical processes in the condensed phase and open the door to many exciting future opportunities.
\end{abstract}

\maketitle

The Born Oppenheimer approximation to separate the electronic and nuclear wavefunctions underpins the concept of potential energy surfaces and forms the bedrock of any modern chemistry course. Much less attention, however, is generally given to the routinely assumed additional approximation employed in atomistic simulations that the nuclear motion and sampling on the resulting electronic energy surface can be treated classically. 
Within the classical nuclei approximation, one loses the ability to describe nuclear zero-point energy, quantization of energy levels, and tunneling, as well as exchange and coherence effects. However, even at room temperature the zero-point energy of a typical chemical bond of frequency $\omega$ (${\sim}\hbar \omega/2$) exceeds the thermal energy scale of that coordinate at temperature $T$ (${\sim}k_{B}T$) by an order of magnitude. These effects can thus make large changes to the structure and dynamics in processes ranging from proton delocalization and tunneling in enzymes \cite{Klinman2013,Sutcliffe2002,wang+14pnas,Wang2016} to changes in the stability of crystal polymorphs \cite{ross+16prl} to the the phase diagram of high pressure melts \cite{mora+13prl}. A revealing consequence of neglecting nuclear quantum effects (NQEs) is that equilibrium isotope effects would be predicted to be zero, despite forming the basis of vital analysis methods in fields ranging from the atmospheric sciences to biochemistry and materials science.

In addition to the importance of calculating and understanding these properties, modelling the quantum nature of the nuclei has become increasingly important due to the greater availability of accurate and affordable methods to describe the electronic potential energy surface on which the nuclei evolve. 
The accuracy of these surfaces is constantly improving, and the most recent generation of state-of-the art potential energy surfaces are now usually generated either by on-the-fly evaluation of the electronic structure, or by fitting it to complex functional forms~\cite{Fanourgakis2008,behl-parr07prl,bart+10prl,Wang2011,Babin2013}, yielding the bare Born-Oppenheimer electronic surface. These high quality descriptions of the interactions are now allowing researchers to go beyond previous empirical surfaces fit to experimental data that have traditionally obfuscated the role of NQEs in these systems by parameterizing them, at a particular state point, into the potential energy surface itself. This approach belies the true physical origins of NQEs which arise from the behavior of the nuclei on the electronic potential surface and not from the surface itself.

The path integral approach to quantum mechanics provides an elegant route to treat NQEs for equilibrium and dynamical properties. This formalism arises from recognizing that the fundamental equations of quantum mechanics can be obtained from a generalization of the minimum action principle of classical mechanics. In the Lagrangian formulation of classical mechanics, a particle always follows the path which minimizes its action. The path integral formulation considers what happens if one instead allows deviations from the classical path. By doing this, Feynman was able to show that by summing the amplitudes associated with all the possible paths quantum mechanics emerges in a physically insightful way~\cite{Feynman1948,feyn-hibb65book}.
Classical mechanics emerges in the limit where all the paths that deviate from the one of minimum action cancel each other due to their positive and negative amplitudes. The path integral formalism thus provides one of the most natural routes to consider building on classical mechanics for the calculation of dynamical properties. 

Unfortunately, the oscillatory amplitudes mean that a na\"ive  implementation of the path integral formalism is extremely expensive - if the problem is classical, one would evaluate all possible paths but classical mechanics would only emerge from the cancellation of all but one path.
Hence, to avoid much of this unnecessary effort, semiclassical and quantum classical methods frequently consider only the paths which deviate a small amount from the classical one and have shown success in a large number of problems, albeit often at significant computational cost.

\begin{figure*}
\includegraphics[width=0.7\textwidth]{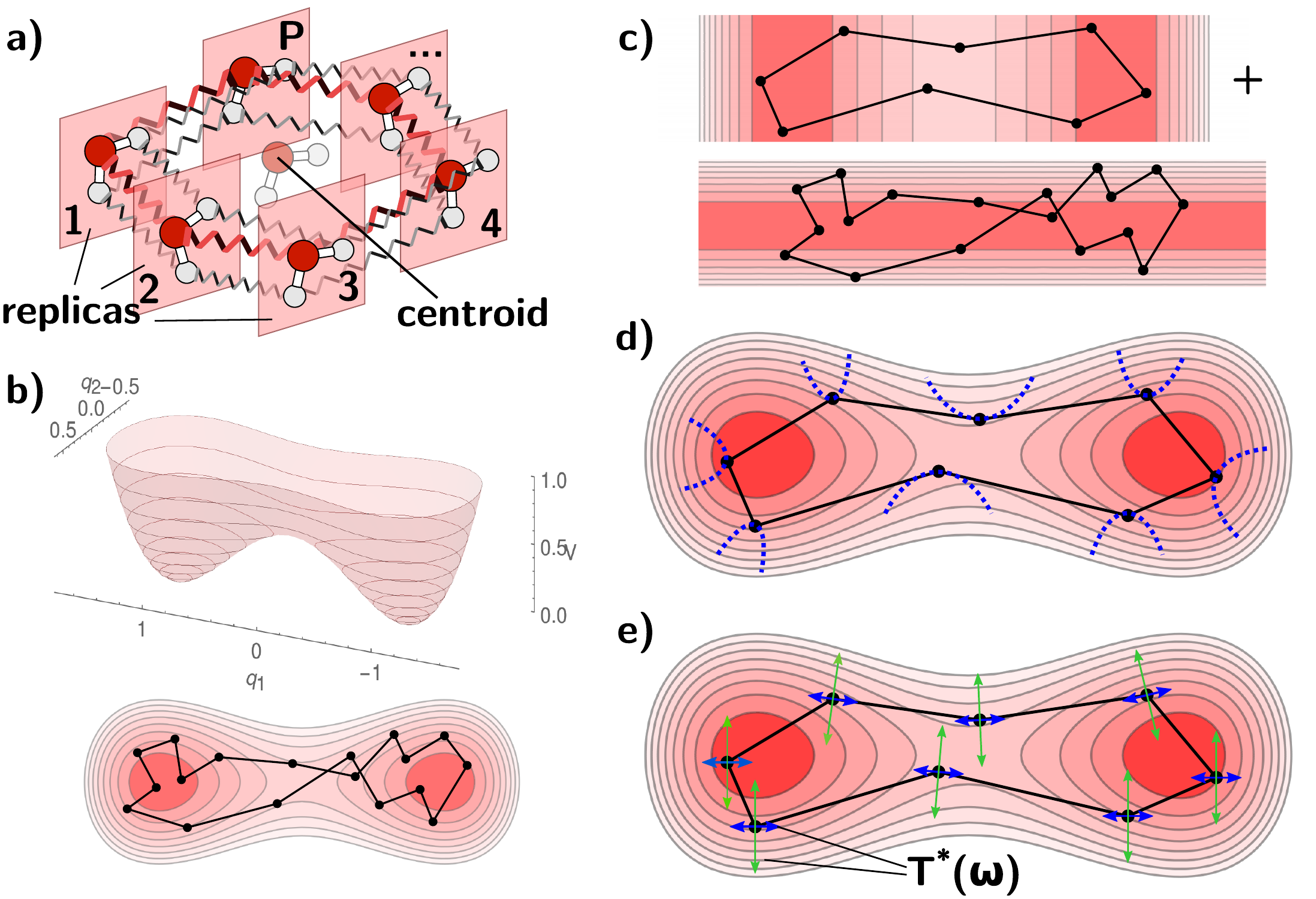}
\caption{\label{fig:intro} (a) Schematic representation of a ring-polymer (imaginary time path integral) simulation of a water molecule. Each replica (bead) comprises physical interaction between one oxygen (O) and two hydrogen (H) atoms, and corresponding atoms in adjacent replicas are 
joined by a harmonic spring. (b) Representative equilibrium configuration of a ring polymer on a two-dimensional potential energy surface composed of a harmonic term along $y$ and a double-well quartic term along $x$. (c) In a ring-polymer contraction scheme, the slowly varying component is evaluated on a contracted version of the ring-polymer, whereas the stiff harmonic term is evaluated on
all replicas. (d) A high-order factorization achieves more rapid convergence with the number of replicas $P$ as it exploits information on the local curvature of the physical potential to achieve more accurate path integration. (e) A generalized Langevin equation can be used to accelerate
the convergence of a PI calculation by selectively and automatically enhancing the fluctuations of the high-frequency vibrations.}
\end{figure*}

However, when the path integral approach is applied to the calculation of static properties, such as free energies and structures, for distinguishable nuclei, the oscillatory amplitudes of the paths that are present when calculating dynamical observables no longer appear. This simplification arises since, to evaluate static properties, one only needs to weight the states of the system by the Boltzmann factor ($e^{-\beta \hat{H}}$). As explained in {BOX 1}, this is typically referred to as the imaginary time propagator, since it is related to the real time propagator ($e^{-\mathrm{i} \hat{H}t/\hbar}$) by the replacement $t = \mathrm{i}\beta\hbar$. When the path integral approach is applied to the imaginary time propagator it allows the calculation of static properties to be exactly mapped onto a classical problem composed of a number of replicas, $P$, of the classical system, with corresponding atoms connected by harmonic springs (Fig.\ref{fig:intro}a).
This object is referred to as the imaginary time path or colloquially as a ``ring polymer''. Introducing this isomorphic representation makes simulating the quantum mechanical system equivalent to performing classical molecular dynamics, although for an extended and hence more complicated system~\cite{parr-rahm84jcp,chan-woly81jcp}.

The imaginary time path integral formalism thus provides a highly appealing approach to exactly include NQEs in static properties and approximate them for the quantum dynamics of chemical systems. In addition, imaginary-time PIMD is also used as the basis for a number of successful approaches to obtain approximate quantum dynamics ranging from quantum transition state theories \cite{Gillan1987a,Gillan1987,Voth1989,Mills1997,Thompson1999,rich-alth09jcp,Althorpe2011,Hele2013,Richardson2016,Hele2016} to the centroid molecular dynamics (CMD)\cite{cao-voth93jcp,cao-voth94jcp,Jang1999}  and ring polymer molecular dynamics (RPMD) methods \cite{crai-mano04jcp,habe+13arpc} and their variants. 

Despite the importance of NQEs, the adoption of path integral methods has traditionally been inhibited both by their computational cost and limited availability in mainstream atomistic simulation software.
In particular, standard PIMD simulations involve a computational cost which is tens to hundreds times of that associated with a classical treatment of the nuclei. This hefty overhead is due to the need to simulate many replicas of the system, each of which involves an additional force evaluation, which typically constitutes the bottleneck of modern atomistic simulations. 
The number $P$ of replicas needed to converge PIMD to the quantum limit, and hence its cost relative to the corresponding classical simulation, is a small multiple of the ratio of the quantum harmonic energy level spacing to the thermal energy, $\hbar\omega_\text{max}/k_{B}T$, that quantifies the ``quantumness'' of the system (see BOX 1). Unfortunately, $P$ grows rapidly as the maximum vibrational frequency $\omega_\text{max}$ in the system increases or as the temperate is decreased. For a typical system containing O--H covalent bonds at room temperature, this leads to at least $P=32$ replicas being required to converge simple structural properties, with larger values required for other properties such as the heat capacity, which require convergence of the fluctuations of the energy.

In the last decade many of these challenges have been addressed by a stream of methodological advances that have dramatically reduced the computational cost of including NQEs, in many cases making it comparable to that of treating the nuclei classically.  
In this review we summarize the concepts underlying these new techniques, and present examples that demonstrate how modelling chemical systems containing hundreds of atoms, accounting fully for nuclear and electronic quantum effects, can now be achieved with modest computational effort. These advances have enabled the shift to a molecular modelling paradigm in which treating light nuclei as quantum particles is increasingly mainstream. 

\section{Accelerating path integral molecular dynamics}

Most of the difficulties associated with the execution of path integral simulations, as well as the ideas to accelerate them, can be understood in terms of the properties of the classical ring polymer that arises from the quantum-classical isomorphism~\cite{chan-woly81jcp}. 
Within the path integral formulation of quantum statistical mechanics (see BOX 1), the quantum mechanical partition function $Z=\operatorname{tr} [ e^{-\beta \hat{H}}]$ (where $\beta=1/k_{B}T$ is the inverse temperature and $\hat{H}$ is the Hamiltonian operator of the system) is mapped onto the classical partition function $Z_P \sim \int e^{-\frac{\beta}{P} H_P}$, corresponding to the Hamiltonian
\begin{equation}
H_P  = \sum_{j=1}^P \frac{ \mathbf{p}_j^2 }{2m} + V(\mathbf{q}_j) + \frac{1}{2}m \omega_P^2 (\mathbf{q}_j-\mathbf{q}_{j+1})^2. \label{eq:trotterham}
\end{equation}
This so-called ring polymer Hamiltonian represents $P$ copies of the physical system, for each of which the potential $V(\mathbf{q})$ must be computed. Adjacent replicas are connected by harmonic springs of frequency $\omega_P=P/\beta\hbar$, with cyclic boundary conditions $j+P\equiv j$ (see Fig.~\ref{fig:intro}a). Since each replica of the system is just an independent classical realization of the chemical system and the only interaction between the replicas arises from the computationally cheap harmonic springs, the cost of evaluating the ring polymer potential energy thus grows linearly with $P$. Position-dependent observables can be obtained by simply evaluating them at the position $\mathbf{q}$ of any of the replicas.
Momentum-dependent observables, require the use of more complicated estimators that depend on replica-replica correlations, which has led to considerable effort being devoted to the development of more efficient forms~\cite{cao-bern89jcp,yama05jcp,ceri-mark13jcp,chen-ceri14jcp,kara-vani17jcp}.
This issue arises because the momenta $\mathbf{p}$ appearing in Eq.~\ref{eq:trotterham} do not correspond to the physical momenta but are simply a sampling device, with no explicit physical meaning, although techniques such as CMD and RPMD attempt to use them to construct approximations to quantum time correlation functions to obtain dynamical properties.

Sampling the classical ring polymer Hamiltonian (Eq.~\ref{eq:trotterham}) is considerably simpler than solving the original quantum mechanical problem for high-dimensional, many particles problems. In addition, since the ring polymer Hamiltonian is just a classical Hamiltonian in an extended phase phase, the plethora of techniques that have been developed to accelerate the sampling, integration and thermostatting of classical molecular dynamics simulations can usually be easily adapted for use in PIMD simulations. However, there are many evident and hidden challenges to be faced. Most obviously, evaluating $H_P$ is $P$ times more demanding than evaluating the classical ($P=1$) Hamiltonian. Convergence is determined by the highest-frequency physical vibrations, and the asymptotic convergence rate falls slowly, as $\mathcal{O}(P^{-2})$. Furthermore, from the last term in Eq.~\ref{eq:trotterham} one can see that the ring polymer Hamiltonian contains frequencies of the order of $\omega_P$, which is, at convergence, many times higher than the maximum physical frequency. These fast, highly harmonic, ring-polymer modes have generally been considered a hard sampling problem, that call for reduced integration time step and aggressive thermostatting. 
In fact, the quasi-harmonic nature of the high-frequency vibrations -- both physical and stemming from the spring term in Eq.~\ref{eq:trotterham} -- and the fact that they are the limiting factor to the convergence of quantum observable underpin all of the most recent and successful approaches to reduce the overhead of path integral methods.

\subsection{Efficient integrators and thermostats for PI}

Efficiently integrating the path integral Hamiltonian in Eq.~\ref{eq:trotterham} presents two main challenges arising from the high frequencies of order $\omega_{P}$ that appear in it. First, these frequencies are typically much higher than the maximum physical frequency, limiting the time step that can be used. Second, the highest of these harmonic frequencies are spectrally well separated from those of the physical system, which makes energy exchange between them highly inefficient, leading to significant sampling and ergodicity problems. However, both of these issues can be addressed so that integration can be performed for path integral simulations with time steps that are usually the same as those used in the corresponding classical simulation.

To improve the size of the integration time step, one can exploit the fact that the free ring polymer Hamiltonian, i.e. without $V(\mathbf{q}_j)$, can be transformed into the normal mode or staging representations~\cite{Sprik1985,tuck+93jcp,tuck+96jcp}. These transformations decouple the adjacent replicas from each other and hence in both of these representations the free ring polymer Hamiltonian, which includes the high frequency ring polymer modes, can be integrated analytically, allowing large time-steps to be used \cite{ceri+10jcp}. In addition, since the momenta in the kinetic energy term in Eq.~\ref{eq:trotterham} are introduced purely for sampling, one can also mitigate the integration issues even further by modifying them to bring down the frequency of the highest ring polymer modes \cite{tuck+93jcp,tuck+96jcp}. 

To achieve efficient sampling and avoid ergodicity issues, one can recognize that in the normal mode or staging representations of the free ring polymer, the frequencies of each mode are known analytically and hence can be targeted with optimally coupled thermostats. This has led to deterministic schemes based on Nose-Hoover chains~\cite{tuck+93jcp,tuck+96jcp} and stochastic schemes using either targeted white noise or colored noise to optimally sample these modes~\cite{ceri+10jcp}. Once these modes are optimally coupled, one is left with just the classical problem of how to efficiently thermostat the diffusive centroid degrees of freedom. Together these approaches, combined with new integrators to give additional stability\cite{liu+16jcp,Mouhat2017}, have alleviated most of the issues that have previously limited time steps and efficiency in path integral simulations.

\subsection{Ring polymer contraction}

Ring polymer contraction (RPC) provides a framework for reducing the cost of evaluating the potential energy and forces on the ring polymer \cite{Markland2008}. In contrast to the methods discussed in the following sections, RPC does not reduce the total number of replicas, $P$, but instead provides a contracted representation of the imaginary time path consisting of $P'$ replicas on which the computationally costly part of the potential energy can be evaluated. RPC exploits the fact that the strong springs between the replicas keep them spatially close and hence that any smoothly varying interaction can be approximated with negligible error on a much smoother representation of the imaginary time path with fewer replicas (see Fig.~\ref{fig:intro}b and c)\cite{Markland2008,Markland2008b,Fanourgakis2009}. In particular, a contraction scheme is characterized by defining a transformation matrix that takes the full $P$-replica description and maps it to a $P'$-replica one~\cite{Markland2008}. The original, and by far the most commonly employed contraction scheme \cite{Markland2008,Markland2008b,Fanourgakis2009,Marsalek2016,Marsalek2017,kapi+16jcp,John2016}, involves transforming to the normal mode representation of the free ring polymer and discarding the $P-P'$ highest normal modes and then transforming back to the Cartesian representation. More recently other procedures such as averaging contraction \cite{Geng2015} and stride contraction\cite{Cheng2016b} have occasionally also been used, although the latter leads to unstable dynamics in some cases. Once the forces have been evaluated on the contracted $P'$-replica ring polymer they are projected back onto the full $P$-replica ring polymer, and combined with other force components before propagating the ring polymer dynamics.

To benefit from RPC, a reference system is required that approximates the rapidly varying forces present in the system. This reference system is subtracted from the full system's forces to leave a smoothly varying difference force which can be evaluated on the $P'$-replica contracted ring polymer. If this reference system is chosen such that its cost is negligible compared to that of the smoothly varying forces, one can decrease the cost of the force evaluations by a factor of $P/P'$. However, it is vital to note that the reference force only has to leave a slowly varying remainder. In fact, the reference force can be something that would give a very poor result for the dynamics and structure of the system if used alone (without the difference force which corrects for its deficiencies). Early applications of RPC with empirical potentials employed reference systems which involved splitting the inter- and intramolecular forces \cite{Markland2008}, range separation of the Coulomb potential \cite{Markland2008b} and in polarizable force fields by splitting of the contributions to the polarization \cite{Fanourgakis2009}. More recently, RPC has been applied to systems with ab initio potential energy surfaces \cite{Geng2015,Marsalek2016,Marsalek2017,kapi+16jcp,John2016}, in a number of cases obtaining dramatic speed-ups \cite{Marsalek2016,Marsalek2017}.

The use of a reference system has natural origins in the multiple times scale (MTS) molecular dynamics methodology, where the reversible reference systems propagator algorithm (r-RESPA)\cite{Tuckerman1992} was formulated as a method to allow efficient propagation in molecular dynamics simulations with multiple components of the force which vary on different time scales. Whereas MTS schemes exploit the slowly varying nature of some forces in real time to extend the propagation time step, RPC takes advantage of the spatially smooth variation of the forces in the imaginary time path integral. The considerations needed for a good reference force in the two approaches are thus similar and so it is natural to utilize MTS and RPC simultaneously\cite{Marsalek2016,kapi+16jcp,Marsalek2017}. RPC is a highly appealing approach as it also allows one to calculate approximate dynamics within the CMD or RPMD frameworks and can further be combined with acceleration approaches that also reduce the overall $P$, such as those discussed below. 

\subsection{High-order PI}

The slow convergence of PIMD with the number of replicas $P$
is a consequence of the fact that the kinetic $K$ and potential energy $V$ terms in the
high-temperature Boltzmann operator $e^{-\beta \hat{H}/P}$ do not commute. Because of this the commonly adopted Trotter factorization (see BOX 1)
introduces an error, which is second-order in the path discretization $\beta\hbar/P$. 
Several more accurate factorizations have been proposed that include corrections
that depend on the commutator $[\hat{V},[\hat{T},\hat{V}]]$. 
This factorized form of $\operatorname{Tr} e^{-\beta \hat{H}}$
can also be mapped onto a classical sampling problem for $P$ replicas, that 
converge to the quantum limit with a leading-order error $\mathcal{O}(P^{-4})$.
However, the corresponding higher-order Hamiltonian~\cite{suzu95pla,chin97pla,taka-imad84jpsj} 
contains a term proportional to $\left|V'\right|^2$. The associated forces,
which are needed in a PIMD scheme, involve the second derivatives
of the potential. This higher-derivatives information is crucial to obtain a more 
effective path integration (see Fig.~\ref{fig:intro}d), but its calculation is impractical 
for all but the simplest potential energy models. To circumvent this problem, most high-order PIMD
simulations have used reweighting schemes~\cite{jang-voth01jcp,yama05jcp,pere-tuck11jcp,mars+14jctc},
that first perform a conventional path integral trajectory, and then weight 
different snapshots with a factor that depends on the exponential of the difference
between the Trotter and the high-order Hamiltonians. Unfortunately, due to the exponential form, the weights vary wildly and introduce statistical inefficiency that is exacerbated for large-scale systems~\cite{ceri+12prsa}. 
Recent solutions to this problem include using a truncated cumulant expansion
of the weighted average~\cite{polt-tkat16cs}, that avoids statistical problems and has
been shown to be affected by remarkably low systematic errors. 
Alternatively, one can borrow some ideas from path integral Monte Carlo 
calculations~\cite{buch-vani13cpl}, and develop a finite-difference scheme
to evaluate the troublesome second derivatives of the potential. It then becomes possible to sample explicitly
high-order path integral Hamiltonians with molecular dynamics~\cite{kapi+16jcp2},
with only a marginal overhead relative to a conventional PIMD simulation with the same $P$. 
High-order schemes are particularly attractive for simulations below room temperature, or whenever
an accuracy in quantum (free) energies of a few meV/atom is required. 

\subsection{Colored-noise methods}

All of the approaches we discussed so far focus on efficiently performing Boltzmann sampling of the classical ring-polymer Hamiltonian. A distinctively different approach, 
instead, builds on the intuition that quantum mechanical fluctuations 
can be effectively mimicked by breaking the classical fluctuation-dissipation theorem. 
In the 1980s, Ford, Kac and others developed a ``quantum Langevin 
equation''~\cite{ford-kac87jsp,ford+88pra} to model the coupling 
between a system and a quantum mechanical bath. More recently, several
groups have proposed to use a Generalized Langevin Equation (GLE)
to enforce the frequency-dependent effective temperature
$T^\star(\omega)=\hbar\omega/2k_B \ \coth \beta\hbar\omega/2$, that mimics 
the effect of quantizing the nuclear degrees of freedom~\cite{buyu+08pre,ceri+09prl2,damm+09prl}. 
Some of these methods are virtually exact when studying the
thermodynamic properties of perfectly harmonic systems. Unfortunately,
in the presence of anharmonic couplings, energy flows between 
high-frequency (hot) and low-frequency (cold) modes, which results
in significant deviations from the desired $T^\star(\omega)$~\cite{ceri+09prl2}. 
While this zero-point energy leakage can be controlled~\cite{ceri+10jctc} 
and some of the dynamical disturbances induced by the GLE corrected~\cite{ross+18jcp},
it would be desirable to extend the method in such a way that it can be systematically converged. 

Such convergence can be achieved using path integral + GLE (PI+GLE) techniques. The general idea of these methods is to design an effective $T^\star_P(\omega)$ such that a $P$-replicas PIMD simulation would give converged results in the harmonic limit for any value of $P$~\cite{ceri+11jcp}.
It should be stressed that $T^\star_P(\omega)$ is designed to reproduce 
the marginal distribution of individual replicas, which is enough to 
accelerate convergence for any structural observable, but not to converge
some of the more exotic estimators that depend on the overall distribution
of the path. While it is possible to include further constraints on the 
distribution, e.g. to accelerate convergence of an estimator of the 
kinetic energy (PIGLET method~\cite{ceri-mano12prl}), one should keep in mind the fact that $T^\star_P(\omega)$ may not be sufficient to converge all estimators when
computing more complex properties, such as equilibrium isotope fractionation ratios~\cite{ceri-mark13jcp}. GLE methods can be applied to any empirical or ab initio potential and combined with the other accelerated techniques. They have been shown to provide a dramatic acceleration in the convergence, up to 100-fold when applied at cryogenic temperatures~\cite{uhl+16jcp}, and can typically model aqueous systems at room temperature with as few as 6 replicas. 

In conclusion, a large tool-kit of methods to accelerate path integral simulations now exist, each of which possess its own benefits and pitfalls. To provide further guidance BOX 2 summarizes each method's strengths and weaknesses, and provides some practical advice to choose the best combination for a given modelling scenario. Finally, it is also worth stressing that many of these methods can be used simultaneously to accelerate path integral simulations even further. 

\section{Applications}

The combination of increased computational power and algorithmic developments over the past decade, discussed in the previous section, have opened the door to apply imaginary time path integral simulations to a wide range of systems. In particular, these advances have facilitated the use of path integrals with ab initio descriptions of the potential energy surface, allowing reactive processes to be studied in systems in areas ranging from biology to materials science. Here we will primarily focus on these recent applications. However, the application of imaginary time path integrals to chemical systems has a long and rich history of pioneering developments and applications~\cite{Tuckerman1997,Marx1998,marx+99nat} spanning a period of over 40 years. For these we refer the reader to earlier reviews \cite{Berne1986,Rossky1991,cepe95rmp,Marx1999Rotors,MarxHutter,marx06cpc,Paesani2009,habe+13arpc,Ceriotti2016}.

\subsection{Aqueous and Biological Systems}

Hydrogen bonded systems and those involving proton networks have been the target of a large number of path integral studies \cite{Ceriotti2016}. Due to the important role of the proton, which can vary from only being slightly shared to fully delocalized between the hydrogen bond donor and the acceptor depending on the strength of the hydrogen bond, quantum effects can play a major role. In recent years it has become clear that NQEs have a somewhat ambivalent impact on the stability of the H bond: quantum fluctuations along the covalent bond permit increased proton sharing and hence strengthen binding, whereas fluctuations in the orthogonal directions facilitate hydrogen bond breaking.~\cite{Habershon2009,Li2011,McKenzie2014} 
The delicate balance between such ``competing quantum effects''\cite{Habershon2009} make accurate PIMD simulations crucial to predict the correct trends, and so the accelerated techniques we discussed in the previous sections have been instrumental to the understanding of these effects in many different systems. In particular, the ability to perform ab initio path integral simulations has allowed the role of competing quantum effects to be investigated in systems that involve reactive events such as proton transfer or delocalization.

One problem where competing quantum effects have been particularly useful is in studies of the fractionation between isotopes. Fractionation of isotopes has recently been investigated in systems encompassing hydrogen/deuterium fractionation between its liquid and vapor phases\cite{Markland2012,Wang2014}, in water and ion clusters \cite{Videla2014,Videla2015} and at the liquid-vapor interface of water \cite{Liu2013} as well as lithium isotope fractionation between aqueous solution and phyllosilicate minerals  \cite{Dupuis2017}. These isotope fractionation ratios, which would be zero if NQEs were neglected, are exploited extensively in applications ranging from monitoring climatic temperature shifts \cite{Zachos2001,Worden2007} to assessing whether low barrier hydrogen bonds are present in biological systems \cite{Harris2000,McKenzie2015}. Indeed, it has recently been shown how one can use the relationship between the difference in the quantum kinetic energy of isotopes in two phases and the fractionation between those phases to provide an approximation of the total quantum kinetic energy for a nucleus in a given chemical phase \cite{chen+16jpcl}. This connection offers an alternative approach to obtaining the absolute quantum kinetic energies of particles requiring only thermodynamic measurements that can be contrasted with the values obtained from deep inelastic neutron scattering experiments \cite{Andreani2005,Pantalei2008}. 

\begin{figure}
\includegraphics[width=1.0\columnwidth]{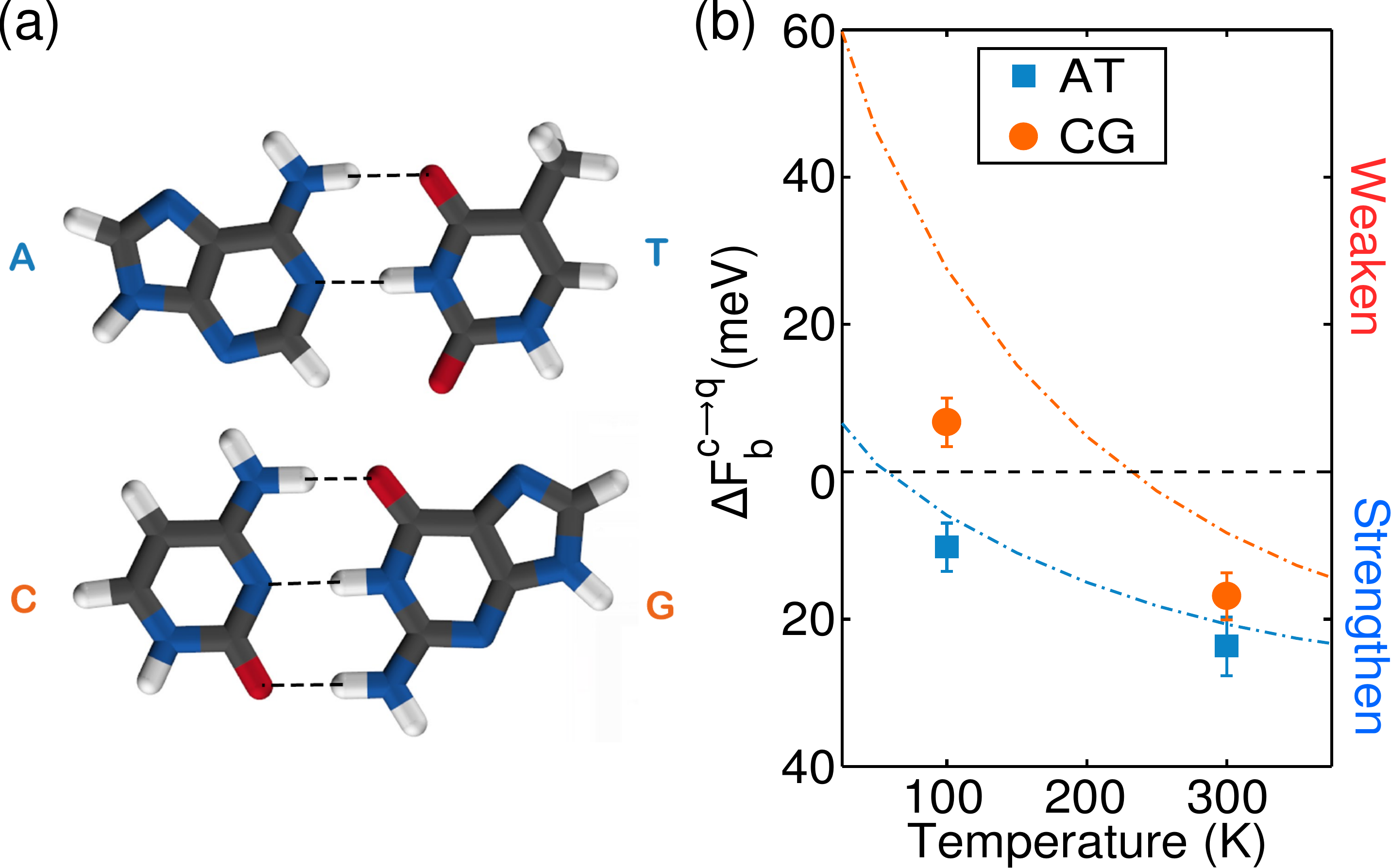}
\caption{(a) Structure of hydrogen-bonded base pair complexes of adenine-thymine (AT) and cytosine-guanine (CG). (b) The binding free energy change due to NQEs as a function of temperature in the AT and CG base pairs obtained from PIMD simulations. Negative values correspond to NQEs strengthening the hydrogen bonds between the base pairs and negative values correspond to weakening. The dashed lines show the harmonic predictions. \emph{Adapted from Ref.~\cite{Fang2016}}\label{fig:dna}}
\end{figure}

A recent ab initio path integral study probed the total free energy change arising from the inclusion of NQEs on the hydrogen bonding of DNA base pairs \cite{Fang2016} building on earlier studies of the role of NQEs in these systems \cite{Perez2010}. Here competing quantum effects give rise to the initially counter-intuitive result shown in Fig.~\ref{fig:dna} that, while at 300~K the hydrogen bonds are strengthened when NQEs are included, when the temperature is lowered to 100~K the strengthening effect decreases to almost zero for all the combinations of base pairs studied. This can be understood in a similar way to liquid-vapor isotope fractionation in water, where the differing temperature dependence of the two competing quantum effects tunes the extent to which they cancel at a given condition, leading to crossovers in the dominant effect \cite{Markland2012,Wang2014}. Many hydrogen bonds fall in the range of O--O donor-acceptor distances around 2.8~\AA~ where the competition between quantum effects is high at 300~K \cite{ross+15jpcl}. However, it is important to note that even in these cases the effect on different properties of the system vary dramatically. For example, in water recent advances have allowed the evaluation of NQEs using potential energy surfaces fit to high level ab initio calculations \cite{Reddy2016} and generated on-the-fly using density functional theory (DFT) \cite{Marsalek2017}. These studies, which give excellent agreement with the structural and dynamical properties of water when NQEs are included, suggest that the NQEs on the diffusion constant and O--O radial distribution function of liquid water are small. However, large changes can still occur in properties that depend sensitively on the proton position. Examples of this include the amount that the proton is shared in the hydrogen bond \cite{ceri+13pnas,Wang2014}, which in turn determines the amount of charge transfer between water molecules and from water molecules to ions upon hydrogen bonding \cite{Schran2017}. Electronic properties, such electronic absorption spectra and electron affinities, can also be highly sensitive to chemical bond lengths and proton positions and hence can exhibit marked NQEs \cite{ceri+13pnas, Hollas2016,Sappati2016}. For instance, GLE methods have been used to asses the quantum effects on the redox properties of small aqueous species in water where large effects ($\sim$0.3~eV) on vertical electron attachment and detachment energies were observed \cite{Rybkin2017}. Hence, even at conditions where one might expect large cancellation due to competing quantum effects on heavy atom properties, the cancellation will not apply equally to all properties, particularly those that depend on the proton positions where the NQEs can still be pronounced. 

\begin{figure}
\includegraphics[width=1.0\columnwidth]{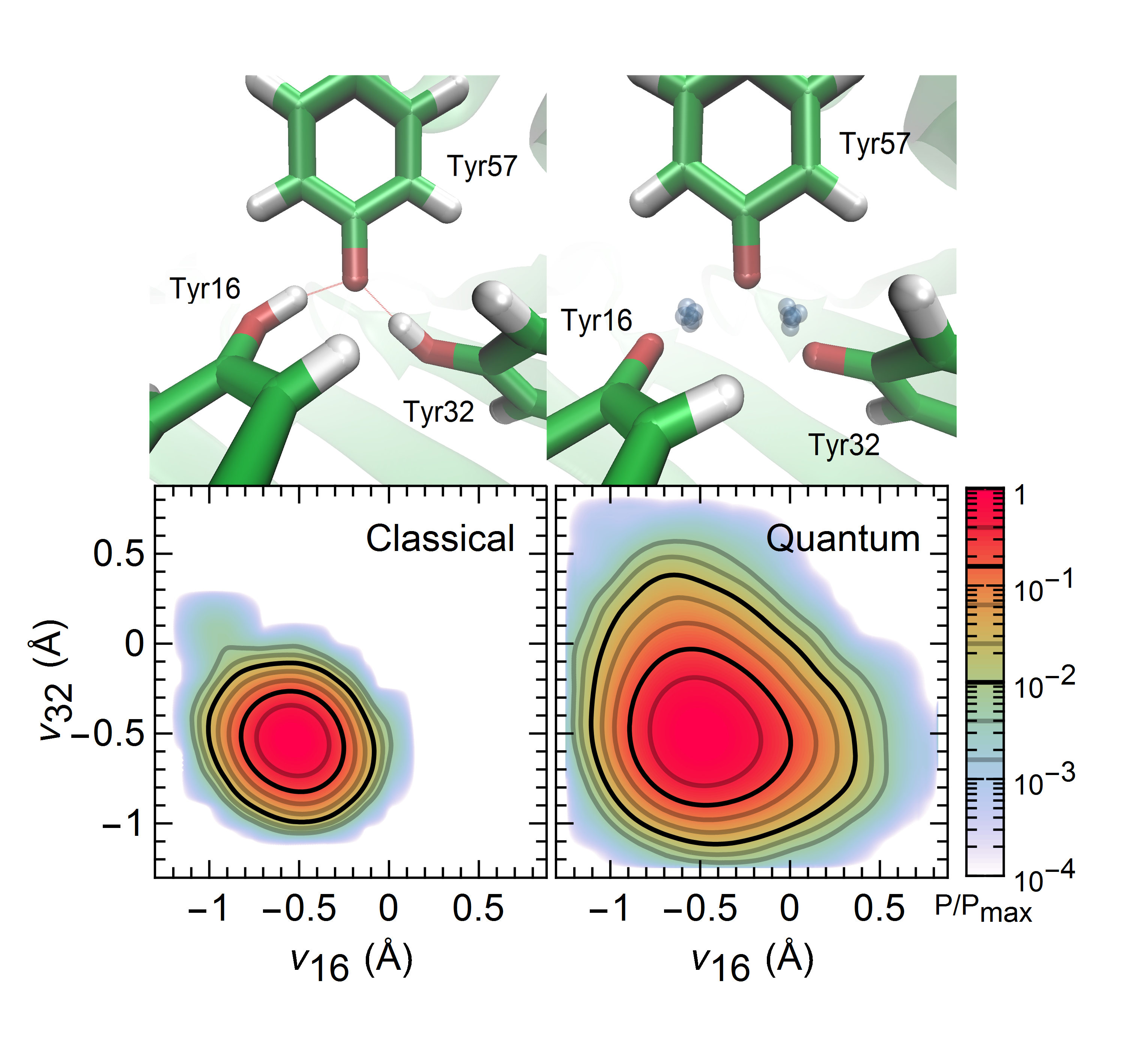}
\caption{Classical and quantum proton sharing distributions in the ketosteroid isomerase enzyme from ab initio molecular dynamics simulations with classical and quantum nuclei. Upon deprotonation of the Tyr57 residue the protons from Tyr16 and Tyr32 can delocalize quantum mechanically to stabilize the residue as shown in the top panels. The bottom panels show the probability distribution along the two proton sharing coordinates, $\nu_{16}$ and $\nu_{32}$ where nuclear quantum effects are shown to markedly increased the sharing. \emph{Adapted from Ref.~\cite{wang+14pnas}}\label{fig:ksi}}
\end{figure}

The balance of the competing effects is tuned by the hydrogen bond strength, the temperature, and the particular chemical property being considered. Biological systems are frequently able to position hydrogen bonds at much shorter distances than observed in liquids, leading them to be further away from the conditions under which NQEs cancel significantly. These low barrier hydrogen bonds, which typically occur when the donor-acceptor distance falls below $\sim$2.6~\AA, are in the regime where the NQEs favoring proton delocalization over disruption is dominant. This effect plays a major role in altering the acidity of the intermediate analog state of ketosteroid isomerase shown in Fig.~\ref{fig:ksi}. In this case, PIMD simulations using GLE acceleration have shown that NQEs cause a 10,000 fold change in the acidity constant of a key tyrosine active site residue by allowing quantum proton delocalization\cite{wang+14pnas,Wang2016} and that this leads to the large active site electric fields~\cite{Wang2017} that have recently been shown to be correlated with its enzymatic efficiency \cite{Fried2014}. 
In addition, the high concentration of strong hydrogen bonds in protein fibrils leads the protons to be extensively delocalized when NQEs are included, and the combination of these two effects has been suggested to be critical in their fluorescence properties \cite{Pinotsi2016}.

While PIMD is now being frequently employed in studies of static properties, the extension of RPC to ab initio molecular dynamics\cite{Marsalek2016,kapi+16jcp} has recently opened the door to the calculation of dynamical properties in condensed phase systems with the approximate CMD and RPMD approaches and their more recent PA-CMD\cite{hone+06jcp} and TRPMD\cite{ross+14jcp} variants (see BOX 3). Until recently the calculation of dynamical properties using these methods was limited to empirical potentials or gas phase molecules. This was due to the combination of the longer time-scales (usually $>$100~ps) required to converge these properties and that, of the acceleration methods mentioned in this review, only RPC is able to provide dynamics that are compatible with the CMD and RPMD methods. Explicit evaluation of the full TRPMD Hamiltonian is possible, and has been recently used to investigate the role of proton transport in water wires \cite{ross+16jpcl}, but is very costly. However, when an appropriate reference potential is available, nanosecond-long path integral simulations on hybrid DFT surfaces for systems of more than 200 atoms are now well within reach \cite{Marsalek2017}.
Recent work has used the latest RPC developments to investigate the role of NQEs on the diffusion, orientational dynamics and IR and Raman spectra of liquid water at 300 K using TRPMD and PACMD \cite{Marsalek2017}. 
This work has revealed that at room temperature NQEs cause a red shift of $\sim$200 cm$^{-1}$ on the OH stretching peak. In addition, it has shown that the accuracy of some simple functionals when used in classical MD is due to a cancellation of errors due to their incorrect treatment of the anharmonicity along the hydrogen bond. More expensive ``hybrid'' functionals are thus needed, and these can only be afforded when the acceleration techniques we discuss here are used to ameliorate their cost.

\subsection{Materials science and matter in extreme conditions}

Computational materials modelling has benefited tremendously by the development of accurate and relatively inexpensive electronic structure calculations that can describe the properties and stability of the most diverse families of compounds without resorting to \emph{ad hoc} empirical potentials~\cite{burk12jcp,klim-mich12jcp,Marzari2016}. Furthermore, interatomic potentials are more and more often built upon high-end electronic structure reference calculations~\cite{behl-parr07prl,bart+10prl} and as such provide a description of the bare Born-Oppenheimer potential energy surface (i.e. those without NQEs parameterized in for a particular state point).
As discussed above, however, there are many properties, from free energies to heat capacities and particle momentum distributions, for which the quantum nature of nuclei can be as important as the underlying description of electronic effects. Indeed, this is often the case for materials containing light elements, at room temperature and below. The availability of methods to treat NQEs more or less inexpensively has made it much easier to probe the qualitative impact they have on complex materials science problems. 

\begin{figure}
\includegraphics[width=1.0\columnwidth]{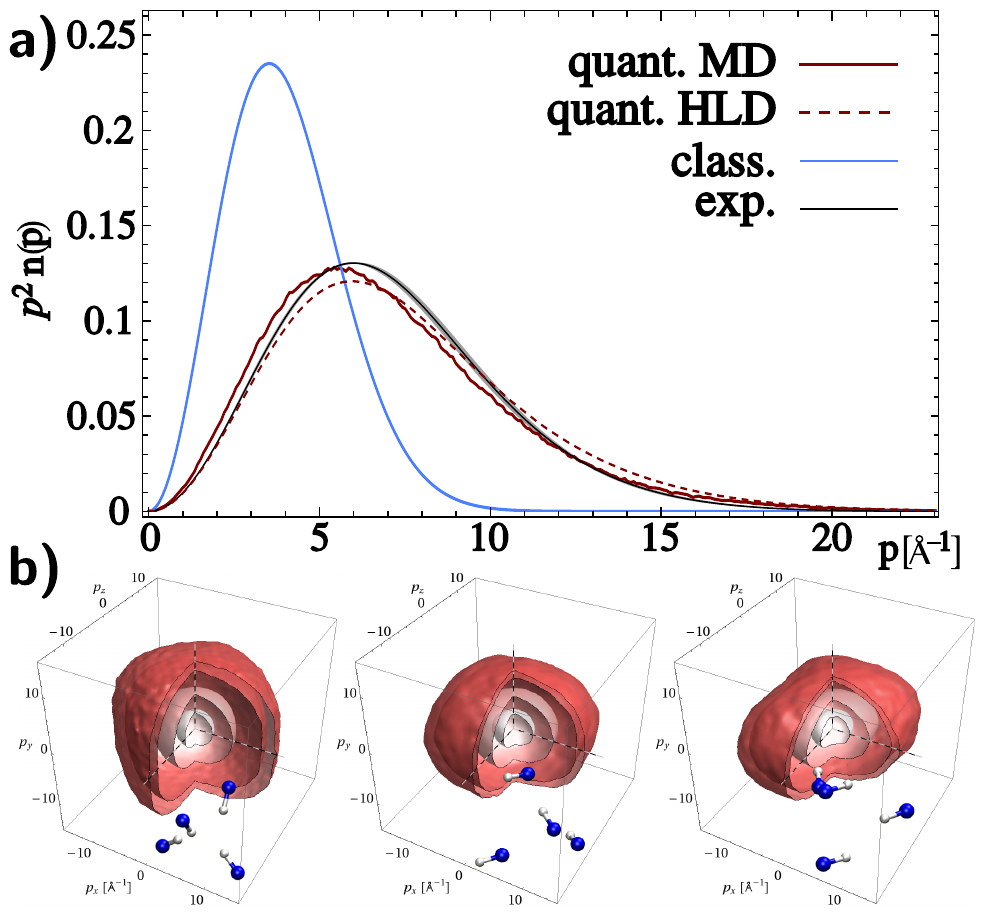}
\caption{(a) Particle-momentum distribution $p^2 n(p)$  for a \ce{Li2NH} polycrystalline sample, as measured by deep inelastic neutron scattering (black line), and as computed using a quantum thermostat (full red line). A Maxwell-Boltzmann distribution at the experimental temperature $T=300K$ (blue line), and the curve computed for a Debye crystal based on the ab initio density of states (dashed red line) are also drawn for reference. (b) The three-dimensional PMD for three proposed crystal structures of \ce{Li2NH}; that of Ref.~\citenum{mice+11prb} (left), that of Ref.~\citenum{muel-cede06prb} (center), and that of Ref~\citenum{magy+06prb} (right) would make it possible to discriminate between the three proposals, that differ mostly by the orientation of the \ce{NH} groups. 
\emph{Adapted from Ref.~\cite{ceri+10prb}}\label{fig:imide}}
\end{figure}

Methods that rely solely on GLEs to include the NQEs have been used particularly often. One of the first real-world applications involved the study of the particle-momentum distribution (PMD) in the hydrogen-storage material \ce{Li2NH}~\cite{ceri+10prb}. 
Despite the inherent limitations of pure GLE methods in terms of accuracy, this example demonstrated that it is possible to capture the deviation of the particle-velocity from a Maxwell-Boltzmann distribution -- an entirely quantum mechanical effect -- achieving semi-quantitative agreement with deep inelastic neutron scattering experiments. 
It is important to note that, despite the quantum thermostat approach\cite{ceri+09prl2,ceri+10jctc} being designed to work in the harmonic limit, it captures some anharmonic effects such as the softening of the high-$p$ tail of the distribution, which can be seen in Figure~\ref{fig:imide}a by contrasting the quantum thermostat results with those obtained from the purely harmonic Debye crystal.

GLEs are also appealing for this application as they make it possible to extract the PMD from direct inspection of the particle velocities (including directionally-resolved information, see Fig.~\ref{fig:imide}b), whereas to obtain this information from a path integral simulation one would have to use open paths, which adds an additional layer of complexity~\cite{morr-car08prl,lin+10prl}.
Other early studies that used colored-noise thermostatting included the determination of the graphite-to-diamond coexistence line, that is bent at room temperature and below due to zero-point fluctuations~\cite{khal+10prb},
and the assessment of the role of quantum delocalization in controlling the balance between Eigen-like and Zundel-like configurations in crystalline \ce{HCl} hydrates, materials that can be seen as a simplified model of the hydrated proton~\cite{hass+12jacs}.
Isotope effects on the lattice dynamics of \ce{LiH} and \ce{LiD} have also been elucidated, using ab initio descriptions of the forces~\cite{Dammak2012}.
More recently, these approaches have been used to study the role of NQEs in problems as complex as the shock-wave compression of fused silica~\cite{Shen2016}, thermal vibrations of carbon nanotubes~\cite{Liu2015}, simulated transmission electron microscopy images~\cite{Lofgren2016},
phase transitions under high-pressure pure~\cite{Bronstein2014}, and salt-doped water ices~\cite{Bronstein2016}.

\begin{figure}
\includegraphics[width=1.0\columnwidth]{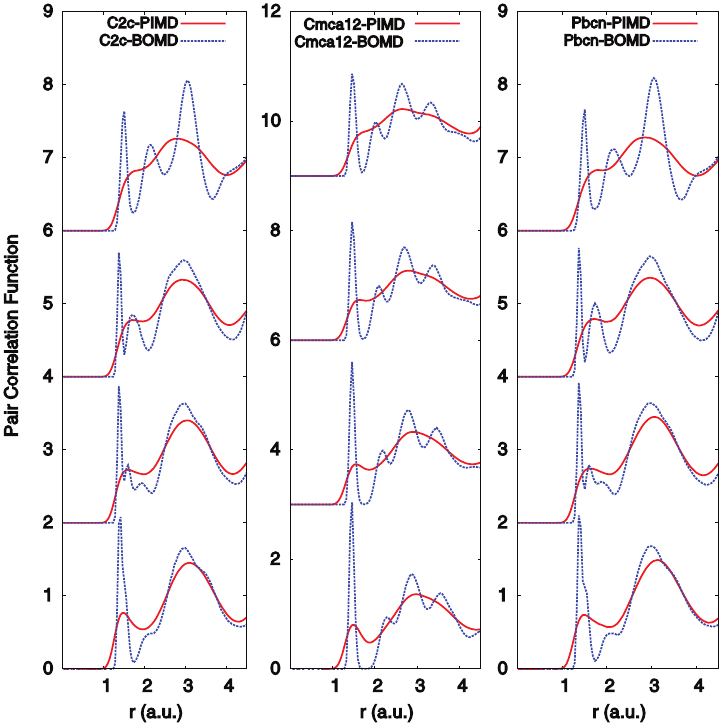}
\caption{Hydrogen-hydrogen radial distribution functions for different phases of solid hydrogen -- $C2c$ (left), $Cmca-12$ (center) and $Pbcn$ (right). All plots correspond to a generalized-gradient approximation density functional, with simulations performed at $T=200$~K. 
Solid red and dashed blue curves correspond to PI+GLE and classical simulations, respectively. Different curves correspond (top to bottom) to pressure values of 350, 300, 250 and 200 GPa.
\emph{Adapted from Ref.~\citenum{mora+13prb}}\label{fig:hydrogen}}
\end{figure}

While GLE based schemes may suffice to assess the qualitative impact of NQEs on the properties of materials, path-integral methods are needed to ensure quantitative accuracy. PI+GLE and PIGLET methods provide the ability to systematically increase the accuracy of the calculation. These methods have made it possible to assess the effect of NQEs in determining the subtle energy balance between different polymorphs of molecular crystals~\cite{ross+16prl}, a class of systems for which free-energy differences of a few tens of meV/molecule suffice to change the stability ranking.
A combination of path integrals and the quantum thermal bath approach has been used to study ferroelectrics, and fuel-cell materials~\cite{brie+16jctc}.

NQEs are particularly important for high-pressure physics, a field that probes the behavior of matter in extreme conditions, such as those found in stars or in giant planets, that are difficult to replicate experimentally.
At GPa pressures and above, the increased confinement of the nuclei means that their quantum mechanical behavior becomes important even well above room temperature.

Examples of simulations probing NQEs under these conditions include the study of the transition between molecular and atomic liquid hydrogen at high pressure and temperature~\cite{mora+13prl}, and the dissociation of water in the GPa pressure regime~\cite{ceri+14cpc}.  
At temperatures around or below room temperature the case for using path integral sampling is particularly compelling. Figure~\ref{fig:hydrogen}, shows the dramatic change in the computed radial distribution functions of different phases of solid hydrogen, using DFT combined with the PI+GLE method~\cite{mora+13prb}. 
Quantum fluctuations substantially smooth out structural correlations, and it is clear that only a simulation that includes them will be able to be in agreement with the experimental structural and thermodynamic properties. The use of colored-noise thermostats combined with PIMD have made it possible to reach an accurate description of these fluctuations with a reasonable computational effort, without having to compromise on the accuracy of the underlying potential energy surface. The possibility of directly using ab initio potentials, and even to couple quantum Monte Carlo and PIMD~\cite{arXiv:1708.07344}, is particularly appealing in applications to exotic states of matter, for which the development of empirical potentials that can span multiple phases is particularly challenging. 

It is also worth mentioning a few recent studies that have used PIMD without utilizing acceleration schemes, as they underscore the urgency of making quantum simulations of nuclear degrees of freedom more affordable. For example, PI simulations have been crucial in determining the impact of NQEs on the delocalization of H atoms and their diffusivity in proton conducting media such as triflic acid hydrates~\cite{haye+09jpcb} and concentrated phosphoric acid~\cite{vilc+12nchem,Heres2016}, as well as in  perovskite oxides~\cite{Zhang2008} and metals such as iron~\cite{kimi+11prb}, and on Ni surfaces~\cite{Suleimanov2012}.
Simulations of crystals in high-pressure conditions, such as the high-temperature superconducting  \ce{SH3} phase of hydrogen sulfide~\cite{erre+15prl,erre+16nature}, is another field in which accessible and accurate modelling of NQEs will increase significantly the predictive power of atomistic simulations.

\section{Outlook and Ongoing Challenges}

The number and variety of recent publications investigating the role of NQEs that we discussed, which is by no means exhaustive, should leave no doubt about the importance of this class of physical phenomena. 
Developing better methods to model NQEs has become particularly urgent, since potential energy surfaces are increasingly obtained (either directly or by statistical learning) from electronic-structure calculations, that yield the bare Born-Oppenheimer potential rather than on empirical force fields that can account for NQEs in an effective manner.
To address this growing need, in the last few years several algorithms have been introduced that reduce the computational effort needed to accurately treat the quantum fluctuations of nuclei. Whereas previously a path integral calculation could easily have been tens to hundreds of times more demanding than a classical molecular dynamics simulations, state-of-the-art methods now curb this cost to barely more than that of a classical simulation. 
One of the remaining challenges limiting the widespread adoption of the toolbox of methods (see BOX 2) to accelerate modelling of NQEs lies in their technical nature, and the fact that they are not yet incorporated into many of the major atomistic simulation codes.
Many of these methods have now been added to commonly used simulations packages~\cite{vand+05cpc,plim95jcp,Eastman2013,Tuckerman2000}. Further, the wrapper code i-PI~\cite{ceri+14cpc} is now available that offers all the classes of acceleration methods discussed in this review, and that can be easily interfaced with any code which can produce a potential energy surface~\cite{vand+05cpc,gian+09jpcm,plim95jcp,SIESTA,AIMS,Aradi2007}. 

However, while the simulation of NQEs for static equilibrium properties of distinguishable particles has now been made affordable for most systems, there are several further directions where methodological advances are needed. One is the combination of the methods reviewed here with path integral Monte Carlo schemes for indistinguishable particles~\cite{cepe95rmp,MartinBook,Walewski2014,Walewski2014b}, which have also benefited greatly from acceleration schemes that cut substantially on the computational expense~\cite{Boninsegni2006}. 
A considerably more challenging open problem involves the simulation of NQEs on the dynamical properties of materials. As discussed in BOX 3, approximate methods inspired by imaginary-time path integrals, such as CMD and RPMD, provide a highly promising approach to compute dynamics incorporating the quantum nature of the nuclei for complex, condensed-phase systems. For this class of techniques, the options to reduce the computational overhead are much more limited, being restricted essentially to RPC (and sometimes rather inaccurate single-bead GLEs for high-frequency spectral properties). What is more, the formal justification of both CMD and RPMD does not involve a hierarchy of well-controlled approximations, making it hard to systematically address their many known artifacts~\cite{Habershon2008,witt+09jcp,ivan+10jcp,ross+14jcp2}. Some success has been recently shown using a (canonical) GLE to improve the quality of vibrational spectra\cite{ross+18jcp}, which leaves some hope that a more principled approach to determine the most suitable form of history-dependent noise may allow the injection of more physics into these approximate approaches. Quantum dynamics, both on a single Born-Oppenheimer surface and also with extensions to non-adiabatic dynamics, is receiving increasing attention \cite{Shushkov2012,Ananth2013,Richardson2013,Kretchmer2016,Shakib2017}, and it is not unreasonable to expect that there may be significant breakthroughs in the near future. However, as far as quantum statistics is concerned, the variety of acceleration techniques that are available and their accessible implementations suggest that the time is ripe for simulations incorporating NQEs to become an essential part of the toolkit of any theoretical chemist, materials scientist, or condensed matter physicist.

\subsection{Acknowledgments}
\acknowledgments
M.C was supported by the European Research Council under the European Union's Horizon 2020 research and innovation programme (grant agreement no. 677013-HBMAP) and the  Swiss National Science Foundation (Project No. 200021-159896). T.E.M was supported by the U.S. Department of Energy, Office of Science, Office of Basic Energy Sciences under Award Number DE-SC0014437 and the National Science Foundation under Grant No. CHE-1652960. T.E.M also acknowledges support from a Cottrell Scholarship from the Research Corporation for Science Advancement and the Camille Dreyfus Teacher-Scholar Awards Program.

\appendix 

\section{Glossary}

\begin{itemize}
    \item {\bf Tunnelling} The ability of quantum particles to pass through a barrier rather than traversing over it, as required in classical mechanics.
    \item {\bf Zero-point energy} The minimum amount of energy a quantum particle must possess even at 0 K.
    \item {\bf Centroid} The center of the imaginary time path that is obtained by taking the mean position of the replicas that comprise it.
    \item {\bf Exchange effects} The effects arising from exchange of indistinguishable particles in quantum mechanics. These are generally significant for electrons, but are often small for nuclei, except at low temperatures.
    \item {\bf Ergodicity} The assumption that as a particle evolves in time it will visit all states with the appropriate frequency associated with the required distribution (e.g. Boltzmann).
    \item {\bf Normal mode and staging representations} Ways of decoupling the spring terms in the imaginary time path integral Hamiltonian.
   \item {\bf Quantum thermostat/quantum thermal bath} Methods to include quantum effects by applying a non-equilibrium Langevin equation to a classical molecular dynamics simulation 
   \item {\bf Estimator} A formula to compute an observed property from simulation data. 
\end{itemize}

\section{BOX 1: The Path Integral Formulation of Quantum Statistical Mechanics}

The calculation of static equilibrium quantum mechanical properties for a finite temperature system of distinguishable particles (as nuclei can be generally considered except at cryogenic temperatures) requires the evaluation of the canonical partition function of the quantum mechanical system,
\begin{equation}
Z = \operatorname{tr} \left[ e^{-\beta \hat{H}} \right] = \sum_{n} \brckt{n}{e^{-\beta \hat{H}}}{n} = \sum_{j} e^{-\beta E_j},
\label{eq:pf}
\end{equation}
where $\{ \ket{n} \}$ is a complete set of states, $\beta=1/k_{B}T$ is the inverse temperature and $\operatorname{tr}$ denotes the quantum mechanical trace over states. The Hamiltonian operator, which here we take to be one dimensional for simplicity, $\hat{H}(\hat{p},\hat{q}) = \hat{T}(\hat{p}) + \hat{V}(\hat{q})$, is the sum of the kinetic and potential energy operators of the system. The third equality is obtained if one evaluates this trace in the basis of the energy eigenstates of the Hamiltonian, where it simply reduces the partition function to a sum over the quantum mechanical energy levels. 

However, calculating the energy eigenstates of the Hamiltonian for the fully interacting condensed phase system of nuclei is computationally intractable. To avoid this problem, one can instead evaluate the trace in the basis of position eigenstates, $\{\ket{q}\}$. Evaluating the trace in a basis of position eigenstates yields,
\begin{equation}
Z = \int dq~ \brckt{q}{ e^{-\beta (\hat{T} + \hat{V})} }{q}. 
\end{equation}
One can then consider making the approximation
\begin{equation}
e^{-\beta (\hat{T} + \hat{V})} \approx  e^{-\beta \hat{V}/2} e^{-\beta \hat{T}} e^{-\beta \hat{V}/2}. 
\label{eq:throwingitallaway}
\end{equation}
This represents a severe approximation with a leading order error term of $\mathcal{O}(\beta^{3})$ as $\hat{T}$ and $\hat{V}$ are quantum mechanical operators that do not commute. Indeed, making the approximation in Eq.~\ref{eq:throwingitallaway} and then proceeding exactly leads to the classical partition function, i.e. the entire quantum mechanical nature of the nuclei is thrown away. The path integral approach provides a route to control this error by instead first splitting the propagator, $e^{-\beta \hat{H}}$, into $P$ parts
\begin{equation}
e^{-\beta \hat{H}} = \left[ e^{-\beta \hat{H}/P} \right]^{P}. 
\end{equation}
A complete set of position states ($\hat{1} = \int dq_{j} \ket{q_{j}}\bra{q_{j}}$) is then inserted between each of these parts of the propagator, yielding matrix elements of the form, $\brckt{q_{j}}{\exp(-\beta \hat{H}/P)}{q_{j+1}}$,
where $j=1,2,\ldots,P$ and the cyclical boundary condition, $j+P \equiv j$, arises due to the trace in Eq.~\ref{eq:pf}. Applying the splitting in Eq.~\ref{eq:throwingitallaway} gives an error that decays as $\mathcal{O}(\left[\beta/P\right]^{3})$ for each of the $P$ matrix elements, yielding a global error that scales as $\mathcal{O}(P^{-2})$ and which therefore converges to the exact result as $P\to\infty$. Exact evaluation of the resulting expression then yields the expression in Eq.~\ref{eq:trotterham}. 

An analogous approach can be used to develop the path integral approach to calculate dynamical properties by instead using the real time propagators of the form $e^{-\mathrm{i}\hat{H}t}$ to time evolve the states. However, due to the complex form of the real time propagator, the paths can possess positive or negative signs, leading to cancellation of paths. This is a manifestation of the dynamical sign problem, which makes the evaluation of the exact real-time properties a highly challenging task.

\section{BOX 3: Approximate quantum dynamics}
The imaginary-time formulation, which provides asymptotically exact results for time-independent equilibrium properties and is the main focus of this review, is also closely connected to several approximate dynamical methods. Imaginary time path integral simulations are used to provide the input for analytic continuation methods\cite{gall-bern96jcp,habe+07jcp}, quantum rate theories\cite{Cao1996,rich-alth09jcp}, and the initial conditions for a wide range range of approximate semiclassical and quantum-classical methods derived from the real time path integral formalism \cite{Feynman1948,feyn-hibb65book}. Furthermore, the CMD\cite{cao-voth93jcp,cao-voth94jcp,Jang1999} and RPMD\cite{crai-mano04jcp,habe+13arpc} approaches use the dynamics generated from the ring polymer Hamiltonian (Eq.~\ref{eq:trotterham}) to directly approximate quantum dynamics. Both techniques can be rationalized heuristically as a classical dynamics on an effective quantum free energy surface, which is sampled exactly, although a number of studies have attempted more rigorous justifications of these methods \cite{Jang1999,Braams2006,Hele2015,Hele2016}. The, perhaps surprising, success of these approaches for many problems can be linked to their preservation of quantum statistics, which dominate the quantum effects in many condensed phase systems due to the rapid quenching of real time quantum coherence effects \cite{Miller2001,Miller2006,Miller2012,Hele2015t,Hele2015}. 

Even though they are based on a similar philosophy, CMD and RPMD differ in three respects. Firstly, CMD prescribes a choice of masses for the beads in the ring polymer Hamiltonian that adiabatically separate the normal modes of the ring polymer from the centroid coordinate. In contrast, RPMD uses the physical masses shown in Eq.~\ref{eq:trotterham}. Second, CMD uses thermostats connected to all the higher normal modes whereas RPMD allows them to evolve freely. Third, CMD evaluates observables on the centroid of the ring polymer while RPMD evaluates them as the average value obtained on each bead. Despite these differences, in many cases CMD and RPMD give remarkably similar results for properties such as diffusion constants, reaction rates and orientational correlation times\cite{ross+14jcp}. 
Significant differences appear, however, for systems at low temperature, particularly when computing properties that are associated with high-frequency spectral features and/or operators that are non-linear functions of the coordinates (e.g. $q^{2}$ vs. $q$), as is the case for many spectroscopic properties. 

In these cases both methods have significant issues, with CMD suffering from the ``curvature'' problem, which leads to spurious red shifting of peaks, while RPMD is plagued by resonance problems, which lead to contamination of the spectrum from the unphysical ring polymer normal modes~\cite{Habershon2008,witt+09jcp,ross+14jcp2}. Due to the similarities of the CMD and RPMD approaches intermediate methods have been recently introduced that seek to balance their inherent problems. These include PACMD~\cite{hone+06jcp}, which retains most of the properties of CMD but does not shift the normal modes to as high frequencies as in full CMD, and TRPMD~\cite{ross+14jcp2}, which largely follows the RPMD prescription but attaches strong thermostats to the normal modes to mitigate the resonance problem~\cite{ross+18jcp}.

\clearpage

\begin{table*}[b]
\caption{
{\bf BOX 2: Pros and Cons of Accelerated PIMD methods}
Given the fast-paced development of new approaches to model NQEs in atomistic simulations, one might feel uncertain as to which method should be chosen for the problem at hand. This table provides some guidelines to provide help in making this choice, by specifying the desirable features each approach does or does not possess. It should also be noted that some of these methods can be applied simultaneously, e.g. RPC/MTS can be combined with both GLE methods and high-order techniques.
}
\begin{tabular} {l | c c c c }
\hline\hline
Features & QT & PI+GLE & high-order & RPC/MTS \\
\hline
Applicable to any potential & + & +  & - (expl.) + (RW, FD, PPI) $^1$ & * needs reference $^2$ \\
Efficient sampling  & - (overdamped) $^3$ & + & - (RW) + (FD, PPI) $^4$ &  + \\
Dynamical properties & * (reconstruction)$^5$ & - & - & * (approximate)$^6$ \\
Well-defined ensemble$^7$ & - & - & + & + \\
Suitable for all estimators & - & * (custom fits)$^8$  & + & + \\
Physically-meaningful $\mathbf{p}$ & * (approximate)$^9$ & - & - & - \\
Typical error in NQEs$^{10}$ & 5-10\% & 1-2\% & $<$1\% & 1-2\% \\
\hline\hline
\end{tabular}\\
\raggedright 
$^1$ Explicit high-order PI require prohibitively expensive evaluation of the second derivatives of the potential. This problem is circumvented by reweighed (RW), perturbed path integrals (PPI) or finite-differences (FD) schemes\\
$^2$ Application of RPC and/or MTS is contingent on the availability of an accurate and inexpensive approximation to the potential \\
$^3$ To compensate for zero-point energy QT simulation require strong damping, that can negatively affect sampling \\
$^4$ RW procedures can lead to large statistical errors, that get worse with system size\\
$^5$ Although dynamics is badly affected by the GLE, it is possible to apply a filtering procedure that recovers meaningful (if approximate) quantum dynamics \\
$^6$ RPC/MTS allows one to run approximate PIMD-inspired quantum dynamics methods such as RPMD or CMD\\
$^7$ The probability of observing a given configuration in a GLE simulation cannot be written as a simple function of the instantaneous configuration, which makes it hard to use techniques (reweighing, replica exchange) that need this term \\
$^8$ In principle it is possible to target specific estimators by reproducing the necessary bead-bead correlations \\
$^9$ Within the approximations of the QT, one can interpret the momentum as a meaningful realization of the quantum mechanical quantum\\
$^{10}$ These order-of magnitude estimates indicate the level of accuracy that can be easily achieved with each method,  based on the Authors' experience for the evaluation of potential or kinetic energy averages in hydrogen-bonded systems.
\end{table*}

\end{document}